\documentstyle[12pt,aasms4]{article}

\slugcomment{Submitted to: The Astrophysical Journal}

\lefthead{Huang et al.}
\righthead{Dynamics of Jetted GRB Remnants}

\begin{document}

\title{Overall Evolution of Jetted Gamma-ray Burst Ejecta}

\author{Y. F. Huang, L. J. Gou, Z. G. Dai, and T. Lu \altaffilmark{1}}
\affil{Department of Astronomy, Nanjing University,
       Nanjing 210093, P. R. China; tlu@nju.edu.cn} 
\authoraddr{Department of Astronomy, Nanjing University,
            Nanjing 210093, P. R. China; 
	    E-mail: tlu@nju.edu.cn } 

\altaffiltext{1}{Also at: China Center of Advanced Science and 
                 Technology (World Laboratory), P.O.Box.8730, Beijing 100080,
		 P. R. China;  Laboratory of Cosmic-Ray and High-Energy 
                 Astrophysics, Institute for High-Energy Physics, Chinese 
                 Academy of Sciences, Beijing 100039, P. R. China }

\begin{abstract}

Whether gamma-ray bursts are highly beamed or not is a very difficult 
but important problem that we are confronted with. Some theorists suggest 
that beaming effect usually leads to a sharp break in the afterglow light 
curve during the ultra-relativistic phase, with the breaking point determined 
by $\gamma = 1 / \theta_0$, where $\gamma$ is the  Lorentz factor of the 
blastwave and $\theta_0$ is the initial half opening angle of the 
ejecta, but numerical studies tend to reject the suggestion. 
We note that previous studies are uniformly based on dynamics that is not 
proper for non-relativistic blastwaves. 
Here we investigate the problem in more detail, paying special attention 
to the transition from the ultra-relativistic phase to the non-relativistic 
phase. Due to some crucial refinements in the dynamics, we can  
follow the overall evolution of a realistic jet till its velocity is as 
small as $\beta c \sim 10^{-3} c$. We find no obvious break in the optical 
light curve during the relativistic phase itself.
However, an obvious break does appear at the transition from the 
relativistic phase to the Newtonian phase if the physical parameters 
involved are properly assumed. Generally speaking, the Newtonian phase 
is characterized by a sharp decay of optical afterglows, with the 
power law timing index $\alpha \sim 1.8$ --- 2.1. This is due to the quick 
lateral expansion at this stage. The quick decay of optical afterglows 
from GRB 970228, 980326, and 980519, and the breaks in the optical 
light curves of GRB 990123 and 990510 may indicate the presence 
of highly collimated $\gamma$-ray burst ejecta.

\end{abstract}


\section{Introduction}

Till the end of August 1999, X-ray, optical, and radio afterglows have been
observed from about 16, 11, and 5 gamma-ray bursts (GRBs) respectively 
(Costa et al. 1997; Bloom et al. 1998; Groot et al. 1998; Kulkarni et 
al. 1998, 1999; Harrison et al. 1999; Stanek et al. 1999; Fruchter et 
al. 1999; Galama et al. 1999a). The so called fireball model (Goodman 
1986; Paczy\'{n}ski 1986; M\'{e}sz\'{a}ros \& Rees 1992; Rees \& 
M\'{e}sz\'{a}ros 1992, 1994; Katz 1994; Sari, Narayan, \& Piran 1996;
Dermer \& Mitman 1999; Dermer, B\"{o}ttcher, \& Chiang 1999, 2000; 
Dermer 2000) is 
strongly favored, which is found successful at explaining the major 
features  of GRB afterglows (M\'{e}sz\'{a}ros \& Rees 1997; Vietri 1997; 
Tavani 1997; Waxman 1997a; Wijers, Rees, \& M\'{e}sz\'{a}ros 1997; Sari 
1997a; Huang et al. 1998a, b, 1999a, b; Dai \& Lu 1998a, b, c; Dermer, 
Chiang, \& B\"{o}ttcher 1999; Dermer, Chiang, \& Mitman 2000).  However, 
we are still far from resolving the puzzle of GRBs, because their 
``inner engines'' are well hidden from direct afterglow observations.

To unveil the nature of the ``inner engine'', we first need to know the 
energetics involved in a typical burst, which itself depends on two 
factors: (i) the distance scale of GRBs, this has been settled since 
the BeppoSAX discovery of GRB 970228; (ii) the opening angle of 
GRB ejecta, i.e., whether gamma-rays are radiated isotropically or not, 
this question is still largely uncertain. Most GRBs localized by 
BeppoSAX have indicated isotropic energies of $10^{51}$ --- $10^{52}$ ergs, 
well within the energy output from compact stellar objects of solar-mass. 
However, GRB 971214, 980703, 990123, and 990510 have implied isotropic 
gamma-ray releases of $3.0 \times 10^{53}$ ergs (0.17 $M_{\odot} c^2$,
Kulkarni et al. 1998), $1.0 \times 10^{53}$ ergs (0.06 $M_{\odot} c^2$, 
Bloom et al. 1998), $3.4 \times 10^{54}$ ergs (1.9 $M_{\odot} c^2$, 
Kulkarni et al. 1999; Andersen et al. 1999), and $2.9 \times 10^{53}$ ergs 
(0.16 $M_{\odot} c^2$, Harrison et al. 1999) respectively. Moreover, 
if really located at a redshift of $z \geq 5$ as suggested by 
Reichart et al. (1999), GRB 980329 would imply an isotropic gamma-ray 
energy of $5 \times 10^{54}$ ergs (2.79 $M_{\odot} c^2$). Such enormous 
energetics has forced some theorists to deduce that GRB radiation must 
be highly collimated in these cases, with half opening 
angle $\theta \leq 0.2$, so that the intrinsic gamma-ray energy could be 
reduced by a factor of $10^2$ --- $10^3$, and could still come from 
compact stellar objects (Pugliese, Falcke, \& Biermann 1999). 
Obviously, whether GRBs are beamed or not is of 
fundamental importance to our understanding of their nature.

How can we tell a jet from an isotropic fireball? Gruzinov (1999) has 
argued that optical afterglows from a jet can be strongly polarized, 
in principle up to tens of percents, if co-moving magnetic fields 
parallel and perpendicular to the jet have different strengths and 
if we observe at the right time from the right viewing angle. More 
direct clues may come from afterglow light curves. Rhoads (1997, 1999a, 
b) has shown that the lateral expansion (at sound speed) of 
a relativistic jet (with a Lorentz factor $\gamma \geq 2$) will 
cause the blastwave to decelerate more quickly, leading to a sharp 
break in the afterglow light curve.  The power law decay indices 
of afterglows from GRB 980326 and 980519 are anomalously 
large, $\alpha \sim 2.0$ (Groot et al. 1998; Owens et al. 1998), 
and optical light curves of GRB 990123 and 990510 even show obvious 
steepening at observing time $t \geq 1$ --- 2 d (Kulkarni et al. 1999; 
Harrison et al. 1999; Castro-Tirado et al. 1999). 
Recently GRB 970228 was also reported to 
have a large index of $\alpha \sim  1.73$ (Galama et al. 1999b).
These phenomena have been widely regarded as the evidence of the 
presence of relativistic jets (Sari, Piran, \& Halpern 1999; 
Castro-Tirado et al. 1999).  

However, numerical studies of some other authors (Panaitescu \& 
M\'{e}sz\'{a}ros 1998; Moderski, Sikora, \& Bulik 2000) have shown 
that due to the increased swept-up matter and the time delay of the 
large angle emission, the sideway expansion of the jet does not lead 
to an obvious dimming of the afterglow. Thus there are two opposite 
conclusions about the jet effect: the analytical solution predicts a 
sharp break, while the numerical calculation shows no such sharp 
breaks. It is very interesting to note that a recent analytic 
treatment by Wei \& Lu (2000) shows that the sharp break predicted 
by Rhoads is usually not exist unless the beaming angle is very 
small. This analysis seems to have given some supports to the numerical studies. 

We find that previous studies on jet effects need to be improved urgently in the 
following three aspects: (i) Afterglows during the mildly relativistic 
($2 \leq \gamma \leq 5$) and non-relativistic ($\gamma \leq 1.5$) phases 
of the expansion are of great importance to us, since they may correspond 
to observing time of $t \sim$ 2 --- 5 d (Huang et al. 1998a, b). However, 
conventional dynamical model can not transit correctly from the 
ultra-relativistic phase ($\gamma \gg 1$) to the non-relativistic phase for 
adiabatic shocks. This has been stressed by Huang, Dai, \& Lu (1999, 2000). 
Therefore previous numerical 
studies were based on improper dynamical equations. They could describe 
an ultra-relativistic jet, but they gave spurious results in the mildly 
relativistic and non-relativistic phases (Huang, Dai, \& Lu 1999a, b); 
(ii) It is reasonable to assume that the lateral expansion speed of the 
beamed ejecta is just the co-moving sound speed $c_{\rm s}$. Usually we 
take $c_{\rm s} = c/\sqrt{3}$, where $c$ is the speed of light 
(Rhoads 1997, 1999a). However in realistic case we expect $c_{\rm s}$ to 
vary with time, and especially it will by no means be $c/\sqrt{3}$ when 
$\gamma \sim 1$. This is another reason that previous studies are not 
proper for mildly relativistic and non-relativistic jets; (iii) In previous 
studies, the expansion of the beamed ejecta is supposed to be solely 
adiabatic. In fact the blastwave should be highly radiative at first. 
It evolves slowly into an adiabatic blastwave at later 
stages (Dai, Huang, \& Lu 1999).

This paper is aiming to overcome the problems mentioned just above. We 
present our refined dynamical model in Section 2. It can describe the 
overall evolution of a realistic, jetted GRB ejecta. 
Synchrotron radiation from shock 
accelerated electrons is formulated in Section 3. Our detailed numerical 
results are presented in Section 4. Section 5 is a brief discussion.

\section{Dynamics} 

\subsection{Basic Equations}

Let $R$ be the radial coordinate in the burster frame; $t_{\rm b}, 
t_{\rm co}$, and $t$ be the time from the event measured in the burster 
frame, co-moving ejecta frame, and terrestrial observer's frame respectively; 
$\gamma_0$ and $M_{\rm ej}$ be the initial Lorentz factor and ejecta mass 
and $\theta$ the half opening angle of the ejecta. The burst energy is 
$E_0 = \gamma_0 M_{\rm ej} c^2$. 

It is useful to write down the following simple relations at first:
\begin{equation}
\label{drdtb1}
d R = \beta c d t_{\rm b},
\end{equation}
\begin{equation}
\label{dtbdt2}
dt_{\rm b} = \gamma dt_{\rm co} 
	   = \gamma (\gamma + \sqrt{\gamma^2 - 1}) dt,  
\end{equation}
where $\beta = \sqrt{\gamma^2 - 1}/ \gamma$.

The evolution of radius ($R$) and swept-up mass ($m$) is described by
\begin{equation}
\label{drdt3}
\frac{d R}{d t} = \beta c \gamma (\gamma + \sqrt{\gamma^2 - 1}),
\end{equation}
\begin{equation}
\label{dmdr4}
\frac{d m}{d R} = 2 \pi R^2 (1 - \cos \theta) n m_{\rm p},
\end{equation}
where $n$ is number density of surrounding interstellar medium (ISM) 
and $m_{\rm p}$ is mass of proton. The opening angle $\theta$ is 
usually evaluated as $\theta \approx a/R$, where $a$ is co-moving 
lateral radius of the ejecta (Rhoads 1999a; Moderski, Sikora, \& Bulik 2000). 
This is not ideal in numerical evaluations since $\theta$ may be as 
large as 0.5 --- 0.8 at later times. Here we use the following 
differential equation: 
\begin{equation}
\label{dthetadt5}
\frac{d \theta}{d t} \equiv \frac{1}{R} \frac{d a}{d t}
		     = \frac{c_{\rm s} (\gamma + \sqrt{\gamma^2 - 1})}{R}.
\end{equation}

As for the evolution of the Lorentz factor $\gamma$, the following 
equation has been widely used for both isotropic fireballs and beamed 
ejecta:
\begin{equation}
\label{dgdmold6}
\frac{d \gamma}{d m} = - \frac{\gamma^2 - 1} {M},
\end{equation}
where $M$ is the total mass in the co-moving frame, including internal 
energy $U$, $dM = [(1 - \epsilon) \gamma + \epsilon] dm$ (Chiang 
\& Dermer 1999; Piran 1999; Moderski, Sikora, \& Bulik et al. 2000). 
Here $\epsilon$ is the 
radiative efficiency defined as the fraction of the shock-generated 
thermal energy (in the co-moving frame) that is radiated. $\epsilon = 1$ 
corresponds to highly radiative case and $\epsilon = 0$ corresponds to 
adiabatic expansion. However Huang, Dai, \& Lu (1999a, b) have demonstrated 
clearly that for adiabatic expansion equation~(\ref{dgdmold6})
does not agree with the Sedov solution in the non-relativistic limit: 
equation~(\ref{dgdmold6}) gives $\beta \propto R^{-3}$ while the Sedov 
solution requires $\beta \propto R^{-3/2}$ (Sedov 1969).  They have 
proposed a new equation. Now we briefly repeat their derivation.

In the observer's frame, since the total kinetic energy of the fireball is 
$E_{\rm k} = (\gamma -1) (M_{\rm ej} + m) c^2 + (1 - \epsilon) \gamma U$ 
(Panaitescu, M\'{e}sz\'{a}ros, \& Rees 1998), and the radiated thermal energy 
is $\epsilon \gamma (\gamma-1) dm c^2$ (Blandford \& McKee 1976), we have
\begin{equation}
\label{dgu7}
d[(\gamma -1) (M_{\rm ej} + m) c^2 + (1 - \epsilon) \gamma U] 
     = - \epsilon \gamma (\gamma - 1) dm c^2. 
\end{equation}
For the item $U$, it is usually assumed that $dU = (\gamma - 1) dm c^2$ 
(Panaitescu, M\'{e}sz\'{a}ros, \& Rees 1998). 
Equation~(\ref{dgdmold6}) has been derived just 
in this way. Huang, Dai, \& Lu (1999a, b) suggested that 
$U = (\gamma -1) m c^2$ is a better approximation that makes 
equation~(\ref{dgu7}) suitable for both ultra-relativistic and 
Newtonian shocks. Using this expression,  
it is easy to obtain (Huang, Dai, \& Lu 1999a, b)
\begin{equation}
\label{dgdmass8}
\frac{d \gamma}{d m} = - \frac{\gamma^2 - 1}
       {M_{\rm ej} + \epsilon m + 2 ( 1 - \epsilon) \gamma m}. 
\end{equation}
Huang, Dai, \& Lu (1999a, b) have shown that this equation is acceptable 
for both radiative and adiabatic fireballs, and in both ultra-relativistic 
and non-relativistic phases. Here we use this equation to follow the 
overall evolution of beamed GRB ejecta.

Equations~(\ref{drdt3}), (\ref{dmdr4}), (\ref{dthetadt5}), and 
(\ref{dgdmass8}) present a thorough description of jet evolution. But 
before evaluating them numerically, we should give the expression for 
$c_{\rm s}$ and $\epsilon$.

\subsection{Sound Speed}

The lateral expansion is determined by the co-moving sound speed. 
The simple assumption of $c_{\rm s} = c/\sqrt{3}$ is not proper 
for our purpose in this paper. We must derive $c_{\rm s}$ from 
$c_{\rm s}^2 \equiv (dp'/de')_{\rm s}$, where $p'$ and $e'$ are 
co-moving pressure and energy density respectively. Kirk 
\& Duffy (1999) have derived 
\begin{equation}
\label{kirkcs9}
c_{\rm s}^2 = \frac{\hat{\gamma} p'}{\rho '} 
	      \left[ \frac{(\hat{\gamma}-1) \rho '}
			   {(\hat{\gamma}-1) \rho ' + \hat{\gamma} p'} 
  	      \right] c^2 ,
\end{equation}
where $\rho '$ is co-moving mass density, 
and $\hat{\gamma}$ is the adiabatic index. Dai, Huang, \& Lu (1999)  
gave a simple and useful expression for $\hat{\gamma}$, 
$\hat{\gamma} = (4 \gamma + 1) / (3 \gamma)$.  
Since $e' = \gamma \rho ' c^2$ 
and $p' = (\hat{\gamma}-1)(e' - \rho ' c^2)$, 
it is easy to get
\begin{equation}
\label{cssquare10}
c_{\rm s}^2 = \hat{\gamma} (\hat{\gamma} - 1) (\gamma - 1) 
	      \frac{1}{1 + \hat{\gamma}(\gamma - 1)} c^2 . 
\end{equation}

In the ultra-relativistic limit ($\gamma \gg 1, \hat{\gamma} \approx 4/3$), 
equation~(\ref{cssquare10}) gives $c_{\rm s}^2 = c^2/3$; and in the 
non-relativistic limit ($\gamma \sim 1, \hat{\gamma} \approx 5/3$), we 
simply get $c_{\rm s}^2 = 5 \beta^2 c^2/9$. So equation~(\ref{cssquare10}) 
is a reasonable expression and will be used in our model.

\subsection{Radiative Efficiency}

As usual we assume that the magnetic energy density in the co-moving 
frame is a fraction $\xi_{\rm B}^2$ of the total thermal energy density 
(Dai, Huang, \& Lu 1999)
\begin{equation}
\label{bsquare11}
\frac{B'^2}{8 \pi} = \xi_{\rm B}^2  \frac{\hat{\gamma}\gamma + 1}
		    {\hat{\gamma} - 1} (\gamma - 1) n m_{\rm p} c^2, 
\end{equation}
and that the shock accelerated electrons carry a fraction $\xi_{\rm e}$ of 
the proton energy. Here $\xi_{\rm e}$ is a parameter characterizing the 
efficiency of energy transport from protons to electrons. In realistic 
case, it is probable that $\xi_{\rm e}$ may vary with time, but 
little is known about the detailed mechanism. So we will take 
$\xi_{\rm e}$ as a constant throughout this article.
This implies that the minimum Lorentz factor of the 
random motion of electrons in the co-moving frame is 
\begin{equation}
\label{gemin12}
\gamma_{\rm e,min} = \xi_{\rm e} (\gamma - 1) 
		     \frac{m_{\rm p} (p - 2)}{m_{\rm e} (p - 1)} + 1,
\end{equation}
where $m_{\rm e}$ is electron mass and $p$ is the index characterizing 
the power law energy distribution of electrons. We here consider only 
synchrotron emission from these electrons, and neglect the contribution 
of inverse Compton emission because the latter is unimportant particularly 
at late times (Waxman 1997a). The energy of a typical electron is lost 
due to both synchrotron radiation and expansion of the ejecta, 
thus the radiative efficiency of this single electron is given by 
$t^{\prime -1}_{\rm syn}/(t^{\prime -1}_{\rm syn}+t^{\prime -1}_{\rm ex})$
(Dai, Huang, \& Lu 1999), where $t^\prime_{\rm syn}$ is the synchrotron 
cooling time, $t^\prime_{\rm syn}=
6\pi m_{\rm e} c/(\sigma_{\rm T}B^{\prime 2}\gamma_{\rm e,min})$,
with $\sigma_{\rm T}$ the Thompson cross section, 
and $t^\prime_{\rm ex}$ is the co-moving frame expansion time,
$t^\prime_{\rm ex}=R/(\gamma c)$. Since all the shock accelerated electrons 
carry only a fraction $\xi_{\rm e}$ of the internal energy, the radiative
efficiency of the total ejecta can be given by (Dai, Huang, \& Lu 1999)
\begin{equation}
\label{radeps13}
\epsilon=\xi_{\rm e} \frac{t^{\prime -1}_{\rm syn}}{t^{\prime -1}_{\rm syn}
	 +t^{\prime -1}_{\rm ex}} .
\end{equation}

For the highly radiative expansion, 
$\xi_{\rm e}\approx 1$ and $t^{\prime}_{\rm syn} \ll t^\prime_{\rm ex}$, 
we have $\epsilon \approx 1$. The early evolution of the ejecta is likely 
in this regime. For an adiabatic expansion, $\xi_{\rm e} \ll 1$ 
or $t^{\prime}_{\rm syn} \gg t^\prime_{\rm ex}$, we get $\epsilon \approx 0$.
The late evolution is believed to be in this regime. 
So one expects that in realistic case the radiative efficiency evolves 
from about 1 to 0. In this paper we call the jet whose radiative efficiency 
evolves according to equation~(\ref{radeps13}) a ``realistic'' one (Dai, 
Huang, \& Lu 1999). 

\section{Synchrotron Radiation}

\subsection{Electron Distribution}

In the absence of radiation loss, the distribution of the shock accelerated 
electrons behind the blastwave is usually assumed to be a power law function 
of electron energy, 
\begin{equation}
\frac{dN_{\rm e}'}{d\gamma_{\rm e}} \propto \gamma_{\rm e}^{-p},
\,\,\,\,\,\,(\gamma_{\rm e,min}\leq \gamma_{\rm e} \leq\gamma_{\rm e,max}),
\end{equation}
where $\gamma_{\rm e,max}$ is the maximum Lorentz factor, 
$\gamma_{\rm e,max}=10^8(B^{\prime}/1{\rm G})^{-1/2}$ (Dai, Huang, \& Lu 1999), 
and $p$ usually varies between 2 and 3. However, radiation 
loss may play an important role in the process. 
Electrons with different Lorentz
factors have different radiation efficiencies.
Sari, Piran, \& Narayan (1998) have derived an equation for 
the critical electron Lorentz factor, $\gamma_{\rm c}$,
above which synchrotron radiation is significant,
\begin{equation}
\label{gammac14}
\gamma_{\rm c}=\frac{6 \pi m_{\rm e} c}{\sigma_{\rm T} \gamma B'^2 t}.
\end{equation}
Electrons with Lorentz factors below $\gamma_{\rm c}$ are 
adiabatic ones, and electrons above $\gamma_{\rm c}$ are highly radiative.

In the presence of steady injection of electrons accelerated by the shock,
the distribution of radiative electrons becomes another power law
function with an index of $p+1$ (Rybicki \& Lightman 1979), but the
distribution of adiabatic electrons is unchanged. Then the actual
distribution should be given according to the following cases 
(Dai, Huang, \& Lu 1999):

\begin{description}
\item (i) For $\gamma_{\rm c}\leq \gamma_{\rm e,min}$,
\begin{equation}
\label{dnei15}
\frac{dN_{\rm e}'}{d\gamma_{\rm e}}=C_1\gamma_{\rm e}^{-(p+1)}\,, \,\,\,\,\,\,
(\gamma_{\rm e,min}\leq\gamma_{\rm e}\leq \gamma_{\rm e,max})\,,
\end{equation}
\begin{equation}
\label{dneic16}
C_1=\frac{p}{\gamma_{\rm e,min}^{-p}- \gamma_{\rm e,max}^{-p}}N_{\rm ele}\,, 
\end{equation}
where $N_{\rm ele}$ is the total number of radiating electrons involved.

\item (ii) For $\gamma_{\rm e,min} < \gamma_{\rm c} \leq \gamma_{\rm e,max}$,
\begin{equation}
\label{dneii17}
  \frac{dN_{\rm e}'}{d\gamma_{\rm e}} = \left \{
   \begin{array}{ll}
 C_2\gamma_{\rm e}^{-p}\,, \,\,\,\, & (\gamma_{\rm e,min} \leq \gamma_{\rm e}
					      \leq \gamma_{\rm c}), \\
 C_3\gamma_{\rm e}^{-(p+1)}\,, \,\,\,\, & (\gamma_{\rm c}<\gamma_{\rm e}
					      \leq \gamma_{\rm e,max}),
   \end{array}
   \right. 
\end{equation}
where 
\begin{equation}
\label{dneiic18}
 C_2=C_3/\gamma_{\rm c}\,,
\end{equation}
\begin{equation}
\label{dneiicc19}
 C_3=\left[\frac{\gamma_{\rm e,min}^{1-p}-\gamma_{\rm c}^{1-p}}
      {\gamma_{\rm c}(p-1)}+ \frac{\gamma_{\rm c}^{-p}-\gamma_{\rm e,max}^{-p}}
      {p} \right]^{-1}N_{\rm ele} .
\end{equation}

\item (iii) If $\gamma_{\rm c} > \gamma_{\rm e,max}$, then
\begin{equation}
\label{dneiii20}
\frac{dN_{\rm e}'}{d\gamma_{\rm e}}=C_4\gamma_{\rm e}^{-p}, \,\,\,\,\,\,
(\gamma_{\rm e,min}\leq\gamma_{\rm e}\leq\gamma_{\rm e,max}),
\end{equation}
where 
\begin{equation}
\label{dneiiic21}
C_4=\frac{p-1}{\gamma_{\rm e,min}^{1-p}-\gamma_{\rm e,max}^{1-p}} N_{\rm ele}.
\end{equation}
\end{description}

\subsection{Relativistic Transformations}

In the co-moving frame, synchrotron radiation power at frequency $\nu '$ from 
electrons is given by (Rybicki \& Lightman 1979)
\begin{equation}
\label{pnue22}
P'(\nu ') = \frac{\sqrt{3} e^3 B'}{m_{\rm e} c^2} 
	    \int_{\gamma_{\rm e,min}}^{\gamma_{\rm e,max}} 
	    \left( \frac{dN_{\rm e}'}{d\gamma_{\rm e}} \right)
	    F\left(\frac{\nu '}{\nu_{\rm c}'} \right) d\gamma_{\rm e},
\end{equation}
where $e$ is electron charge, 
$\nu_{\rm c}' = 3 \gamma_{\rm e}^2 e B' / (4 \pi m_{\rm e} c)$, and 
\begin{equation}
\label{fx23}
F(x) = x \int_{x}^{+ \infty} K_{5/3}(k) dk,
\end{equation}
with $K_{5/3}(k)$ being the Bessel function. We assume that this power 
is radiated isotropically,
\begin{equation}
\label{dpdwp24}
\frac{d P'(\nu ')}{d \Omega '} = \frac{P'(\nu ')}{4 \pi}.
\end{equation}

Let $\Theta$ be the angle between the velocity of emitting material
and the line of sight and define $\mu = \cos \Theta$, we can derive the angular 
distribution of power in the observer's frame (Rybicki \& Lightman 1979), 
\begin{equation}
\label{dpdw25}
\frac{d P(\nu)}{d \Omega} = \frac{1}{\gamma^3 (1 - \beta \mu)^3}
			    \frac{dP'(\nu ')}{d \Omega '}
			  = \frac{1}{\gamma^3 (1 - \beta \mu)^3}
			    \frac{P'(\nu ')}{4 \pi},
\end{equation}
\begin{equation}
\label{tnue26}
\nu = \frac{\nu '}{\gamma (1 - \mu \beta)}. 
\end{equation}
Then the observed flux density at frequency $\nu$ is 
\begin{equation}
\label{snue27}
S_{\nu} = \frac{1}{A} \left( \frac{dP(\nu)}{d \Omega} \frac{A}{D_{\rm L}^2} \right)
	= \frac{1}{\gamma^3 (1 - \beta \mu)^3} \frac{1}{4 \pi D_{\rm L}^2}
          P'\left(\gamma(1 - \mu \beta) \nu \right),
\end{equation}
where $A$ is the area of our detector and $D_{\rm L}$ is the luminosity 
distance.

\subsection{Equal Arrival Time Surfaces}

Photons received by the detector at a particular time $t$ are not emitted 
simultaneously in the burster frame. This effect has been emphasized by a 
number of authors (Waxman 1997b; Sari 1997b; Panaitescu \& M\'{e}sz\'{a}ros 
1998) and may be of great importance to jet radiation 
(Moderski, Sikora, \& Bulik 2000;
Panaitescu \& M\'{e}sz\'{a}ros 1999). In order to calculate observed flux 
densities, we should integrate over the equal arrival time surface 
determined by 
\begin{equation}
\label{eqt28}
t = \int \frac{1 - \beta \mu}{\beta c} dR \equiv {\rm const},
\end{equation}
within the jet boundaries.

\section{Numerical Results}

For convenience, let us define the following initial values or parameters 
as a set of ``standard'' parameters: initial energy per solid angle 
$E_0 / \Omega_0 = 10^{54}$ ergs/$4 \pi$, 
$\gamma_0 = 300$ (i.e., initial ejecta mass per solid angle 
$M_{\rm ej} / \Omega_0 = 0.001867 M_{\odot}/ 4 \pi$), 
$n = 1$ cm$^{-3}$, $\xi_{\rm B}^2 = 0.01$, $p = 2.5$, 
$D_{\rm L} = 1.0 \times 10^6$ kpc, $\xi_{\rm e} =1$, $\theta_0 = 0.2$. 
For simplicity, we first assume that the expansion is completely 
adiabatic all the time (i.e. $\epsilon \equiv 0$, we call it an 
``ideal'' jet, distinguishing it from the ``realistic'' jet defined 
in Section 2.3).

Figure 1 shows the evolution of the Lorentz factor. We see that the 
ultra-relativistic phase lasts only for $\sim 10^5$ s, this is the 
period during which equation~(\ref{dgdmold6}) can be safely 
applied. The mildly relativistic phase lasts from $\sim 10^5$ s 
to $\sim 10^{6.5}$ s. In short, the ejecta will cease to be highly 
relativistic at time $t \sim 10^5$ --- $10^6$ s. This clearly demonstrates 
the necessity to replace equation~(\ref{dgdmold6}) with our improved 
expression of equation~(\ref{dgdmass8}). Figure 1 also indicates
that the transition from the relativistic phase to the non-relativistic 
phase occurs roughly at $t \sim 10^{6.5}$ --- $10^7$ s. Figure 2 illustrates 
the time dependence of shock radius.

\placefigure{fig1}

\placefigure{fig2}

In Figure 3 we present the evolution of the jet opening angle $\theta$. 
During the ultra-relativistic phase ($t \leq 10^5$ s), $\theta$ increases 
only slightly. But at the Newtonian stage ($t \geq 10^7$ s), the increase 
of $\theta$ is very quick. Again this figure indicates that the 
critical point is roughly at $t \sim 10^{6.5}$ s. The usual approximation 
of $\theta \approx a / R$ is not a good approach and our 
equation~(\ref{dthetadt5}) is obviously more reasonable. Figure 4 
illustrates jet evolution on the $y$ -- $z$ plane schematically, where 
$z$-axis is just the symmetry axis of the jet and the lateral expansion 
is {\em approximately} at $y$ direction. Note that the jet is 
non-relativistic when $z \approx 3.2 \times 10^{18}$ cm. Again we 
see the quick lateral expansion at the Newtonian stage. In Figure 5 
we show some examples of equal arrival time surfaces on the 
$y$ -- $z$ plane. Here we assume that the angle between jet symmetry 
axis and the line of sight is $\theta_{\rm obs} = 0$.

\placefigure{fig3}

\placefigure{fig4}

\placefigure{fig5}

Figures 6 and 7 illustrate the time dependence of radiative efficiency 
($\epsilon$) and total kinetic energy ($E_{\rm k}$) for ``realistic'' 
jets respectively. Although the highly radiative phase usually lasts 
for very short period ($t \leq 10^4$ --- $10^5$ s), it dissipates 
the total kinetic energy substantially (up to 90\% in less 
than $10^3$ s). This effect may account for 
the confusing phenomena observed in some GRBs: $\gamma$-ray energy 
released in the initial GRB phase ($E_{\gamma}$) is comparable to 
or even higher than the total kinetic energy inferred from 
afterglow observations (Wijers \& Galama 1999).

\placefigure{fig6}

\placefigure{fig7}

Figure 8 shows the optical light curves, computed for R band observation. 
In this figure, the thick solid line corresponds to an ``ideal'' jet with 
``standard'' parameters, and viewing angle $\theta_{\rm obs} = 0$. For 
comparison, other lines are drawn with only one parameter altered or only 
one condition changed. Please note that at end point of each curve, the 
average electron Lorentz factor is already as small 
as $\gamma_{\rm e} \sim 5$, corresponding to bulk Lorentz 
factor of $\gamma \approx 1.002$, thus completely in the Newtonian regime. 

\placefigure{fig8}

>From Figure 8, we see that in no case could we observe the 
theoretically predicted light curve steepening during the {\em relativistic} 
stage itself (i.e., when $t \leq 10^6$ s), consistent with 
previous numerical studies (Panaitescu 
\& M\'{e}sz\'{a}ros 1999; Moderski, Sikora, \& Bulik 2000).
This has been attributed to the effects 
of equal arrival time surfaces and more and more swept-up material. 
Here we would like to propose another more reasonable explanation. 
Rhoads is correct in saying that the lateral expansion 
begins to take effect when $\gamma \sim 1/\theta_0$. In our calculations 
this occurs at $t \sim 10^{5.5}$ s. However, the blastwave is already 
in its mildly relativistic phase at that moment 
and it will become non-relativistic 
soon after that (i.e., when $t \geq 10^{6.5}$ s, see Figure 1). 
So it is not surprising that we could not see any obvious breaks during 
the relativistic phase, they just do not have time to emerge. 

Fortunately we have pointed out that the lateral expansion is even 
more notable in the non-relativistic  phase, so we expect that a 
break should occur in the light curve between the relativistic stage 
and the non-relativistic stage.  The dash-dotted line in Figure 8 
proves our deduction. This line corresponds to an ``ideal'' jet with 
$\xi_{\rm e} = 0.1$ and with other parameters equal to those in the 
``standard'' set. Here the decay in the relativistic phase is 
$S_{\rm R} \propto t^{-1.61}$, in good agreement with theoretical 
results for isotropic fireballs 
(Wijers, Rees, \& M\'{e}sz\'{a}ros 1997; Sari et al. 1998), and the 
decay in the non-relativistic phase is $S_{\rm R} \propto t^{-2.14}$, 
also consistent with Rhoads' prediction for lateral expansion effect 
(but made only for relativistic jets). 
Previous numerical studies are based on dynamical equations improper 
for non-relativistic blastwaves. They could not correctly reveal the 
break between relativistic and Newtonian stages (Panaitescu \& 
M\'{e}sz\'{a}ras 1999; Moderski, Sikora, \& Bulik 2000).

One may ask why other lines in Figure 8 do not show any breaks. The 
reason is they all correspond to $\xi_{\rm e} = 1$ and their 
peaks appear at relatively late stages. For example, the thick solid line 
peaks at about $10^5$ s. Then we can not see the initial power law 
decay (i.e., $\alpha \sim 1.1$)  
in the relativistic phase and can only see the fast decay of 
$S_{\rm R} \propto t^{-2.14}$ at lage stages. However, we should 
note that although they do not show any obvious breaks, they all 
have a common characteristic: a large timing index of 
$\alpha \sim 1.8$ --- 2.1. We suggest that the sharp decline of 
afterglows from some GRBs (with $\alpha \sim 1.7$ --- 2.0) 
itself may just be the evidence of the presence of a jet 
(Huang, Dai, \& Lu 2000a, b, c). 

Dai \& Lu (1999, 2000) have discussed the optical afterglows from 
isotropic fireballs during the non-relativistic phase extensively. 
The most obvious difference between their results and ours is 
that our timing index at non-relativistic stage is reasonably larger.

\section{Conclusion and Discussion}

In this paper we investigate the detailed dynamical evolution of jets  and 
their afterglows. Our model is simple in form and is easy for numerical 
evaluations. Comparing with previous studies, the model is refined in the 
following aspects: 
\begin{description}
\item (i) Equation~(\ref{dgdmold6}) has been widely used in previous 
studies, however, it is not correct for non-relativistic ejecta 
(Huang, Dai, \& Lu 1999a, b). The dynamics here (i.e., mainly our 
equation~(\ref{dgdmass8}) ) is applicable to both ultra-relativistic and 
Newtonian jets, so we could follow the overall evolution 
(from $\gamma \gg 1$ to $\gamma \sim 1$) of a jet by using a single set of 
differential equations. Numerical results indicate that the ejecta will
cease to be ultra-relativistic $10^5$ --- $10^6$ s after the main GRB. 
We should consider this refinement seriously.

\item (ii) We describe the lateral expansion of jets with a refined sound 
speed expression, which gives reasonable approximations during both 
ultra-relativistic and Newtonian phases (Kirk \& Duffy 1999).

\item (iii) The remnant here is more ``realistic'', i.e., the radiative 
efficiency evolves from 1 (corresponding to highly radiative regime) to 0 
(adiabatic regime), according to our equation~(\ref{radeps13}). 

\item (iv) We use a differential equation to describe the increase of jet 
opening angle $\theta$, not simply using $\theta \equiv a /R$. 
\end{description}
Our model also takes many other important factors into account, for 
example: the equal arrival time surface, the distribution of electrons, 
and the viewing angle.

We find that during the ultra-relativistic phase, the opening angle $\theta$ 
increases only slightly, but at the Newtonian stage it increases quickly 
to $\theta \sim 0.8$ --- 1. The highly radiative 
regime (when $\epsilon \sim 1$) lasts usually for very short period, 
however a substantial fraction of the initial kinetic energy could be 
dissipated in less than $10^2$ --- $10^3$ s. As for the light curves 
of optical afterglows, 
we find no theoretically predicted breaks at the critical points where 
the ejecta transits from $\gamma \geq 1/\theta$ to 
$\gamma \leq 1/\theta$, consistent with previous numerical studies 
(Panaitescu \& M\'{e}sz\'{a}ros 1999; Moderski, Sikora, \& Bulik 2000). 
The reason is that at this critical point the lateral expansion just 
begins to take effect, but its action does not prevail over $\theta_0$ 
completely, and also the relativistic phase is too short for the break 
to appear. However we have shown that the lateral expansion is even 
more striking during the non-relativistic phase, so we expect an 
obvious break between the relativistic stage and the non-relativistic 
stage. This has been proved by our calculations made with 
$\xi_{\rm e} = 0.1$. Other calcuations made with $\xi_{\rm e} = 1$ 
do not show any obvious breaks, but they are all characterized by a 
large timing index. 

We conclude that the most obvious characteristic of beamed eject is 
the quick decline of afterglows at late stages (in fact corresponding 
to the non-relativistic phase), with $\alpha \sim 1.8$ --- 2.1, and 
in some cases, we could even see a sharp break in the light curve 
between the relativistic stage and the non-relativistic stage. We
suggest that the quick decay (with $\alpha \sim$ 1.7 --- 2.0) 
of optical afterglows from some GRBs, such as GRB 970228 ($t \leq 10$ d), 
980326, and 980519, and the breaks in the optical light curves of 
GRB 990123 and 990510, are probably due to beaming effect 
(Huang, Dai, \& Lu 2000a, b, c). 

But we should also keep in mind that other factors 
might lead to a large $\alpha$ as well. For example, the ISM density 
may be a decline function of radius (usually $n \propto R^{-2}$), then 
an isotropic fireball can well explain the quick decay of optical 
afterglows from GRB 970228, 980326, and 990519. Additionally, if an 
isotropic GRB remnant sweeps through a uniform ISM ($n \propto R^0$) 
and an uneven ISM (i.e., $n \propto R^{-2}$) in succession, then an obvious 
break is likely to present in the optical light curve. GRB 990123 and 
990510 may also correspond to this case. The suggestion by Dai \& 
Lu (1999, 2000) that the quick decay of afterglows from GRB 980519 and 
990123 is due to an isotropic Newtonian blastwave in a dense ISM (with 
$n \sim 10^6$ cm$^{-3}$) is also possible. So, the degree of beaming is 
still very difficult to determine now. With the progress in observing 
technique, when much more GRBs are localized, maybe we could also 
infer some hints from statistical researches. However the final 
solution may come from systematic deep optical surveys, which are expected 
to find many faint decaying optical sources if GRBs are highly collimated.
They are afterglows from jetted GRBs whose gamma-ray emission deviates 
the line of sight slightly.

\acknowledgments

This work was partly supported by the National Natural Science 
Foundation of China, grants 19773007 and 19825109, the National 
Climbing Project on Fundamental Researches, and the National 
Project of Fundamental Researches (973 Project).

\clearpage

\figcaption{Evolution of the Lorentz factor ($\gamma$). 
The solid line corresponds to an 
``ideal'' jet with ``standard'' parameters. The dashed line is for ``ideal''
jet with $\theta_0 = 0.1$ and with other parameters being equal to those in  
the ``standard'' set. For the meaning of ``ideal'' and ``standard'', please 
see Section 4 in the main text. 
\label{fig1}}

\figcaption{Evolution of the shock radius ($R$). Parameters and line styles 
are the same as in Fig. 1.
\label{fig2}}

\figcaption{Evolution of the half opening angle ($\theta$). Parameters and 
line styles are the same as in Fig. 1.
\label{fig3}}

\figcaption{Schematic illustration of jet evolution on the $y$ -- $z$ plane. 
$Z$-axis is the symmetry axis of the jet and the lateral expansion is 
{\em approximately} at $y$ direction. Parameters and line styles are the 
same as in Fig. 1.
\label{fig4}}

\figcaption{Schematic illustration of the equal arrival time surfaces on the 
$y$ -- $z$ plane. The jet is ``ideal'' and with ``standard'' parameters. The 
viewing angle is assumed to be $\theta_{\rm obs} = 0$. The dash-dotted lines 
from right to left correspond to spherical ejecta shells at observing time 
of $t = 10^8$ s, $10^7$ s, $10^6$ s, and $10^5$ s respectively, enclosed within 
the jet boundary (dotted lines). The solid lines are corresponding equal 
arrival time surfaces. 
\label{fig5}}

\figcaption{Evolution of the radiative efficiency ($\epsilon$) for a ``realistic''
jet with ``standard'' parameters. 
\label{fig6}}

\figcaption{Time dependence of the total kinetic energy ($E_{\rm k}$) of 
a ``realistic'' jet. Parameters are the same as in Fig. 6.
\label{fig7}}

\figcaption{R band afterglows. $S_{\rm R}$ is in units 
of ergs$\cdot$s$^{-1} \cdot$Hz$^{-1} \cdot$cm$^{-2}$. The thick solid 
line corresponds to an ``ideal'' jet with ``standard'' parameters, 
and viewing angle $\theta_{\rm obs} = 0$. Other lines are drawn with 
one parameter altered or one condition changed: 
the thin solid line corresponds to $\theta_{\rm obs} = 0.3$; the dashed line 
corresponds to $\theta_0 = 0.1$; the dash-dotted line corresponds to 
$\xi_{\rm e} = 0.1$; and the dotted line is for a ``realistic''
jet. 
\label{fig8}}

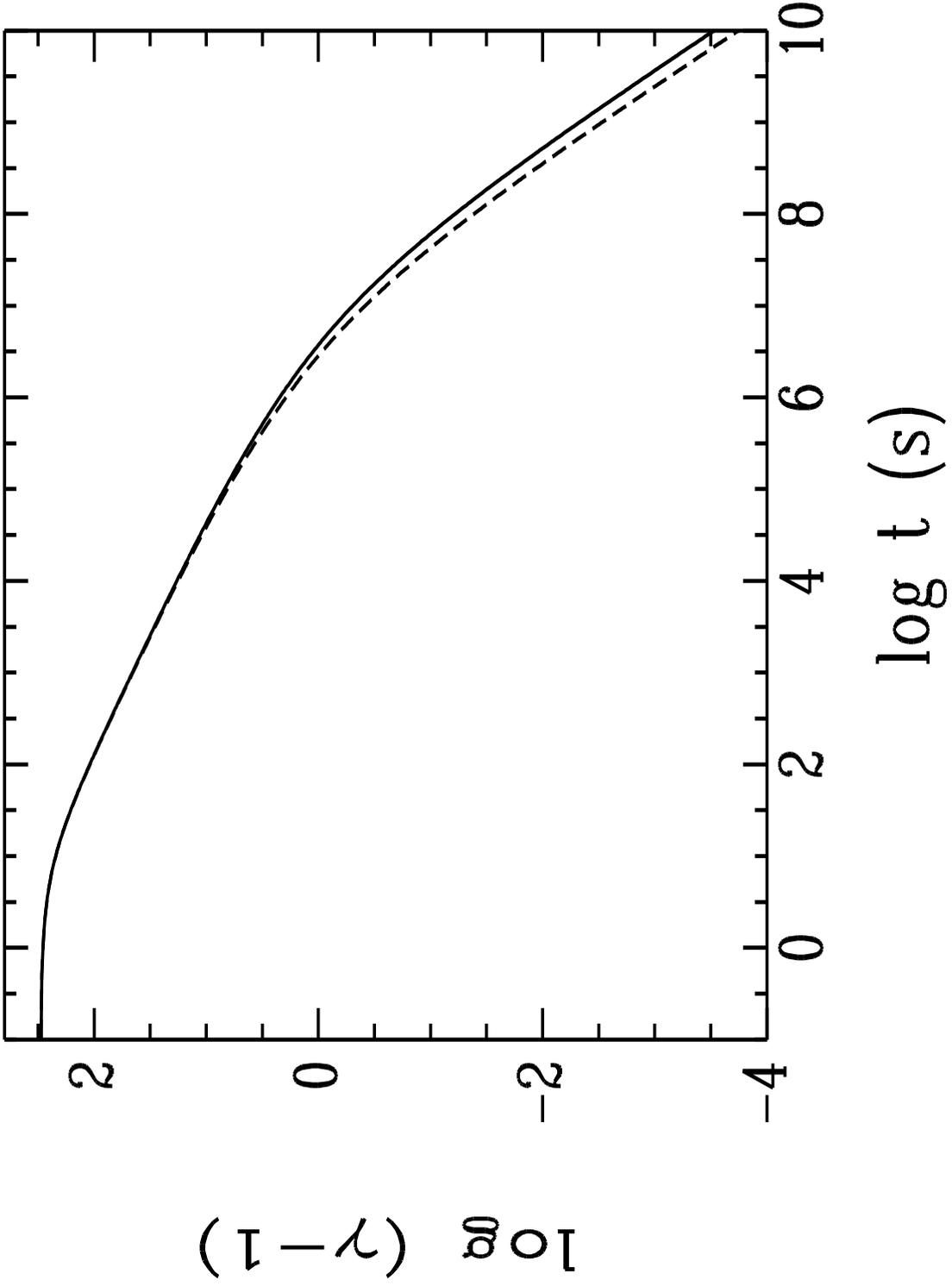
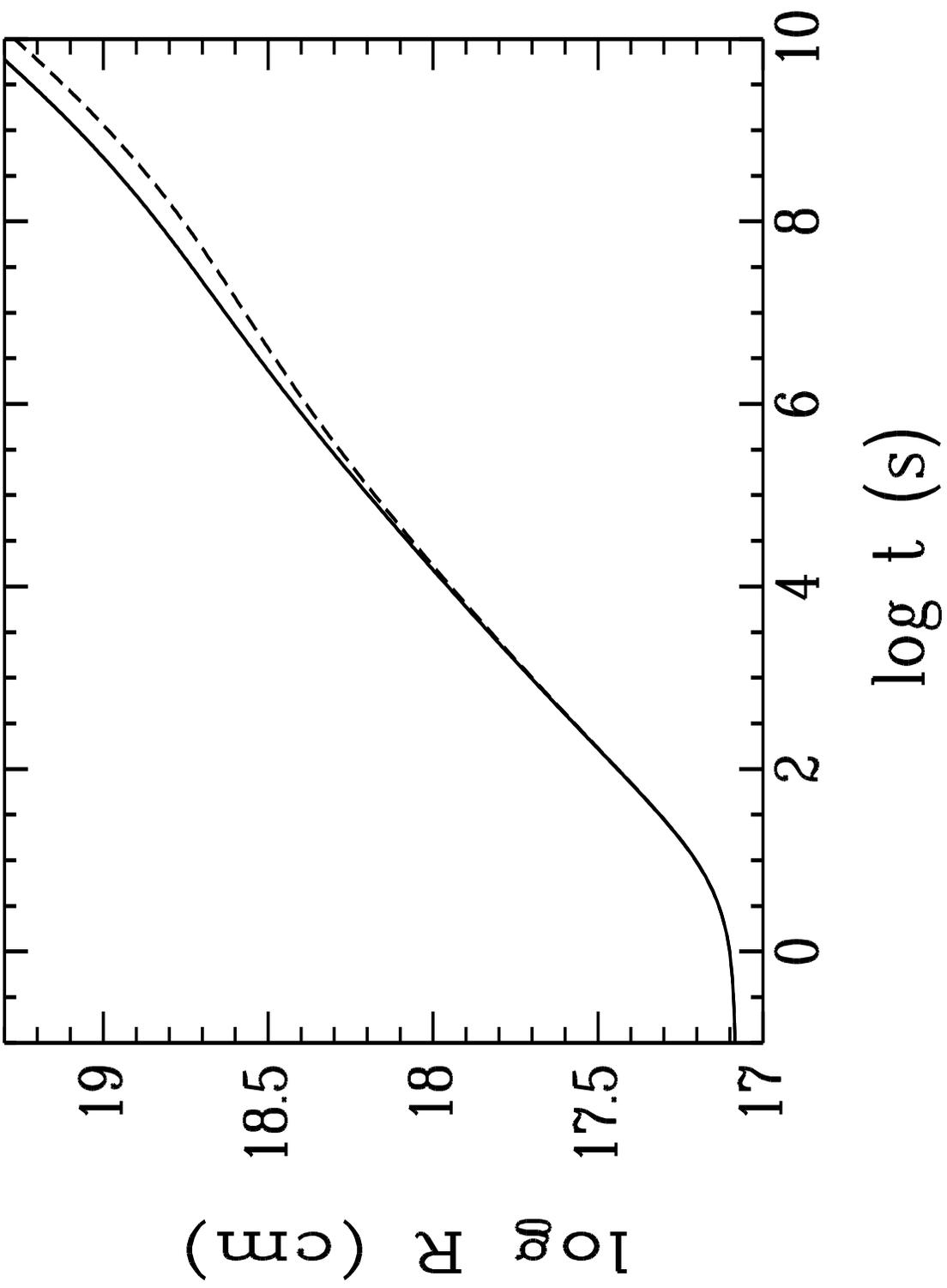
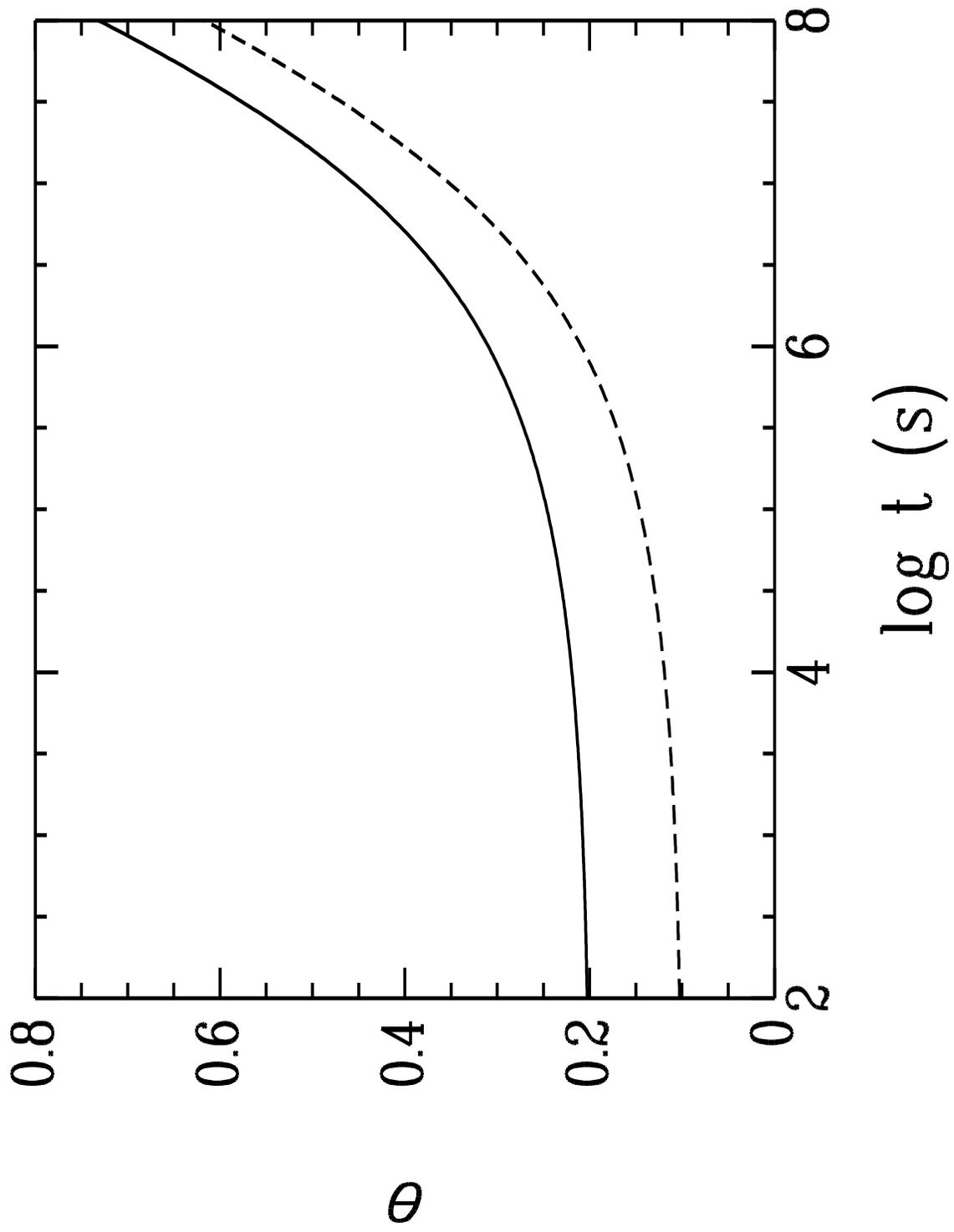
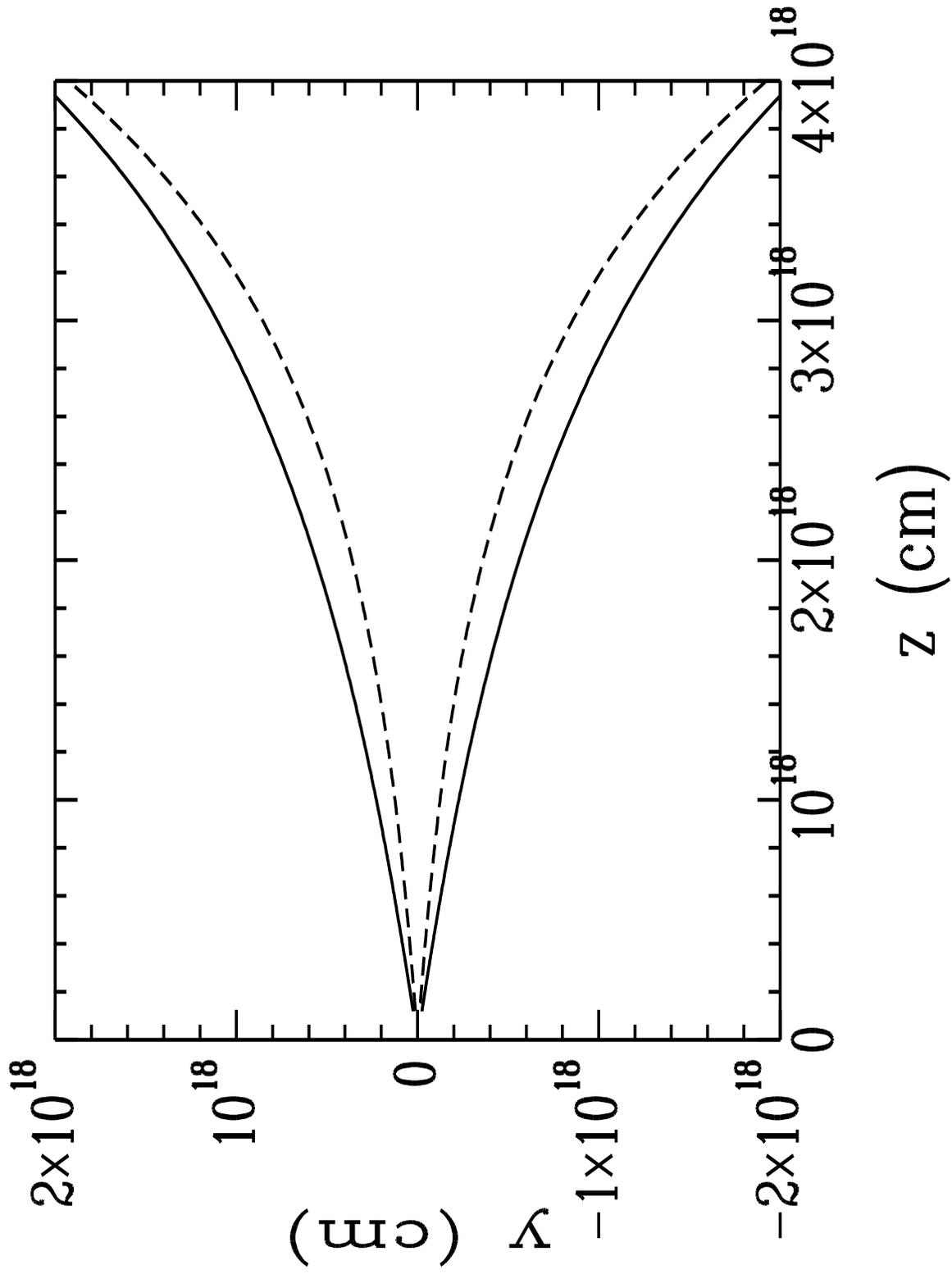
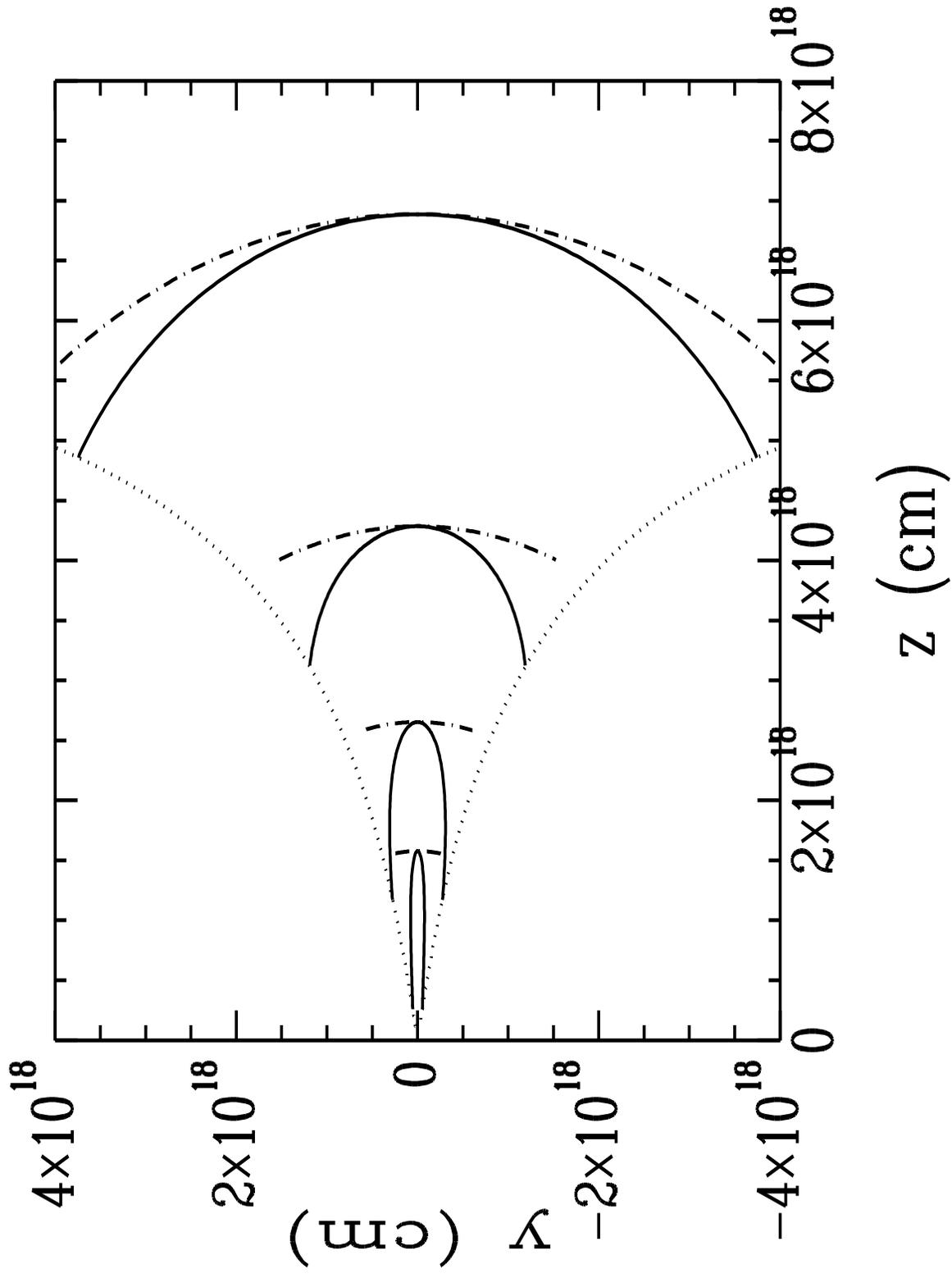
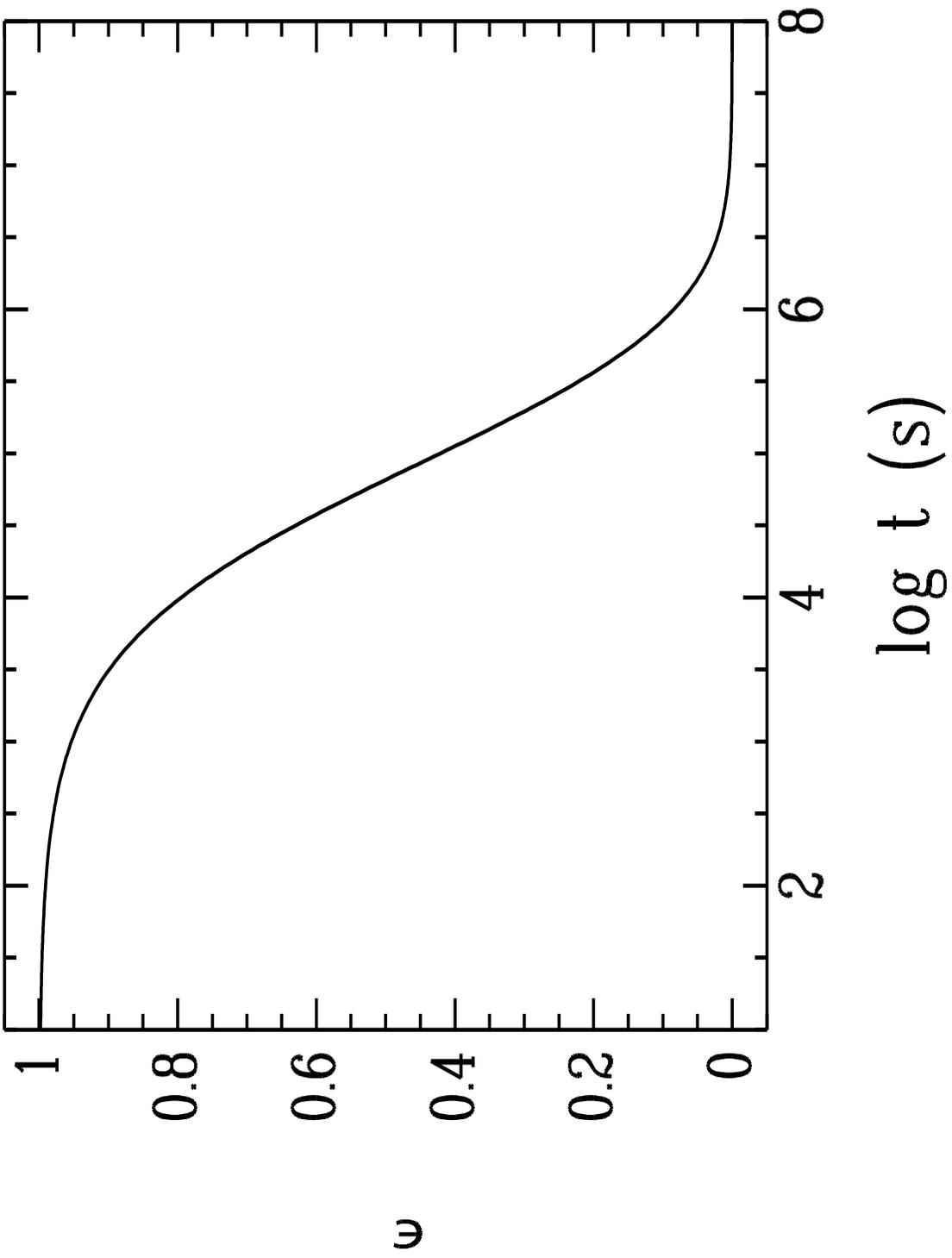
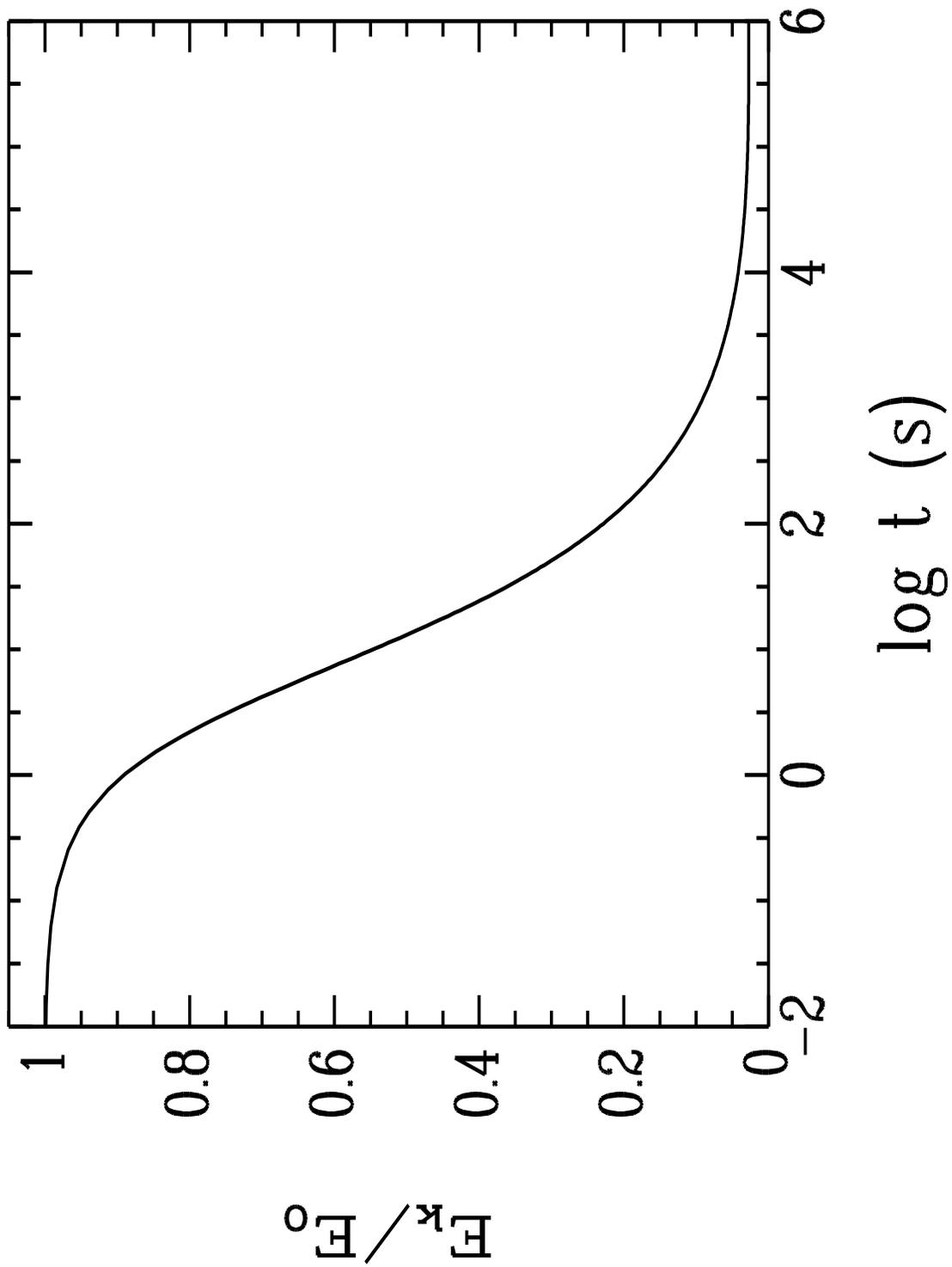
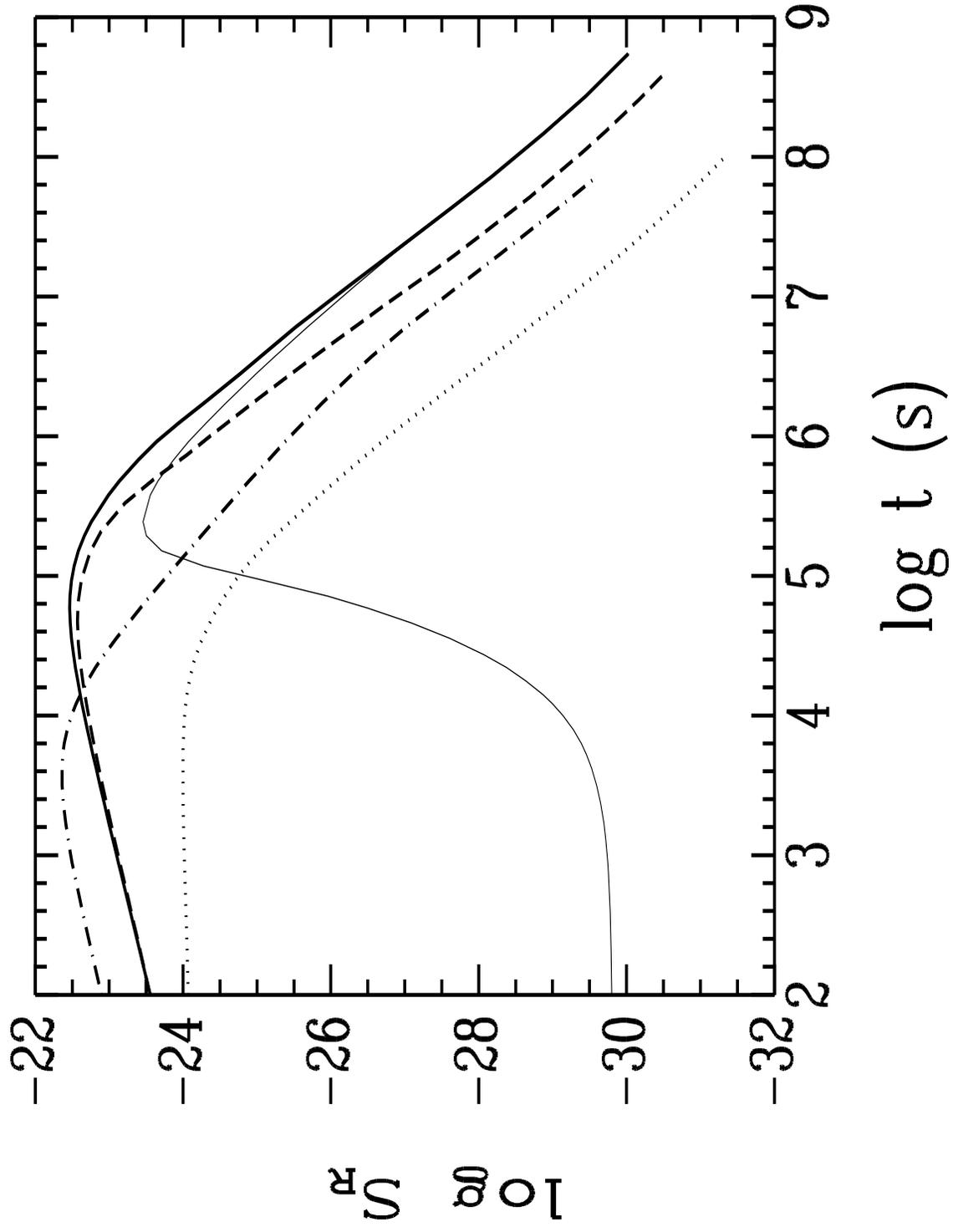
\end{document}